\documentclass[useAMS,usenatbib]{mn2e}

\usepackage{epsfig}

\newcommand{\s}{\,{\rm s}}  \newcommand{\ps}{\,{\rm s}^{-1}}
\newcommand{\yr}{\,{\rm yr}}

\newcommand{\cm}{\,{\rm cm}}
\newcommand{\km}{\,{\rm km}}
\newcommand{\parsec}{\,{\rm pc}}
\newcommand{\kpc}{\,{\rm kpc}}
\newcommand{\erg}{\,{\rm erg}}

\newcommand{\eV}{\,{\rm eV}}

\newcommand{\TeV}{\,{\rm TeV}}
\newcommand{\uG}{\,\mu{\rm G}}
\newcommand{\um}{\,\mu{\rm m}}

\newcommand{\gray}{$\gamma$-ray}
\newcommand{\grays}{$\gamma$-rays}
\newcommand{\twCO}{$^{12}$CO}

\newcommand{\kep}{$K_{\rm ep}$}
\newcommand{\du}{d_{2.5}}
\newcommand{\nH}{n_{\rm H}}
\newcommand{\nt}{n_{\rm t}}
\newcommand{\fo}{f_{\Omega}}
\newcommand{\fc}{f_\mathrm{c}}
\newcommand{\nMC}{n_\mathrm{MC}}

\newcommand{\Chandra}{{\sl Chandra}}

\newcommand{\Fermi}{{\sl Fermi}}
\newcommand{\AKARI}{{\sl AKARI}}

\newcommand{\veritas}{{\sl VERITAS}}
\newcommand{\herschel}{{\sl Herschel}}

\title[The hadronic \grays\ from Tycho's SNR]
{On the hadronic $\gamma$-ray emission from Tycho's Supernova Remnant}

\author[X. Zhang et al.]
{
Xiao Zhang,$^{1}$
~Yang Chen,$^{1,2,}$\thanks{E-mail: ygchen@nju.edu.cn}
~Hui Li$^{1}$~and Xin Zhou$^{3,2,4}$ \\
$^{1}$Department of Astronomy, Nanjing University, Nanjing 210093, P.\ R.\ China \\
$^{2}$Key Laboratory of Modern Astronomy and Astrophysics, Nanjing University, Ministry of Education, China\\
$^{3}$Purple Mountain Observatory, 2 West Beijing Road, Nanjing 210008, China \\
$^{4}$Key Laboratory of Radio Astronomy, Chinese Academy of Sciences, Nanjing 210008, China \\
}

\begin{document}

\date{}

\pagerange{\pageref{firstpage}--\pageref{lastpage}} \pubyear{2012}

\maketitle

\label{firstpage}

\begin{abstract}

Hadronic \gray\ emission from supernova remnants (SNRs) is an important tool to
test shock acceleration of cosmic ray protons. Tycho is one of nearly a dozen
Galactic SNRs which are suggested to emit hadronic \gray\ emission. 
Among them, however, it is the only one in which the hadronic emission is proposed
to arise from the interaction with low-density ($\sim0.3 \cm^{-3}$) ambient medium. 
Here we present an alternative hadronic explanation with a modest conversion 
efficiency (of order 1\%) for this young remnant. With such an efficiency,
a normal electron-proton ratio
(of order $10^{-2}$) is derived from the radio and X-ray synchrotron spectra
and an average ambient density that is at least one-order-of-magnitude higher
is derived from the hadronic \gray\ flux. 
This result is consistent with the multi-band evidence of the presence of dense 
medium from the north to the east of the Tycho SNR. The SNR-cloud association, in
combination with the HI absorption data, helps to constrain the so-far controversial
distance to Tycho and leads to an estimate of 2.5~kpc.

\end{abstract}


\section{Introduction}\label{sect:Intro}

Energetic cosmic rays (CRs) below the ``knee'' energy $\sim 3 \times10^{15}\eV$ 
are usually believed to be accelerated in shock waves of Galactic SNRs through
the diffusive shock acceleration (DSA). Moreover, it has been commonly assumed
that some $10\%$ of the explosion energy of a supernova is transferred to CRs
throughout its whole life in order to explain the observational CR energy density
in our Galaxy \citep[e.g.,][]{Baade1934, Ginzburg1967}.
The theory of DSA of cosmic rays is well established \citep[see e.g.,]
[for reviews, and references therein]{Malkov2001}, but there is no direct evidence for
such acceleration of the CR protons, and the energy conversion efficiency, the 
fraction of the explosion energy converted to CR energy, still remain elusive.

An important clue comes from the hadronic \gray\ emission from the neutral pion 
decay which ensues from the collision of the accelerated protons with the baryons
in proximate dense matter. It is, however, even difficult to distinguish the 
hadronic \gray\ emission from the leptonic emission, and thus the environmental
effect on the \gray\ emission is very noteworthy. So far, \grays\ from some SNRs
have been interpreted to arise from hadronic processes, such as W28, W41, W44, 
W49B, W51C, Cygnus Loop, IC443, CTB~37A, and G349.7+0.2 \citep[see][and references therein]
{Li2012}. The \grays\ from the Cas~A SNR are also explained to have a
hadronic contribution at the GeV energy range at least \citep{Araya2010}. The
hadronic \grays\ from these SNRs are all suggested to be emitted by the interaction
with dense matter. For the \gray\ emission from the Tycho SNR, as a contrast, the
hadronic scenario invokes a tenuous ambient medium \citep*[e.g.,][]
{Giordano2012,Tang2011,Morlino2012}.

Tycho's SNR (G120.1+01.4; 3C~10; SN~1572) is the historical relic of a Type~Ia
supernova as verified with the light echo from the explosion \citep{Krause2008}.
As one of the few accurately known-age SNRs, Tycho's SNR has been well studied over
the entire spectrum. It shows a shell-like morphology in radio with enhanced
emission along the northeastern edge \citep[e.g.,][]{Dickel1991}, and has spectral
index of 0.65 and a flux density of 40.5 Jy at 1.4 GHz \citep{Kothes2006}.
Strong non-thermal X-ray emission is concentrated in the
SNR rim, forming a thin filamentary structure \citep[e.g.,][]{Hwang2002,Bamba2005,Warren2005}.
Recently, TeV and GeV \gray\ emission was successively detected from this SNR
\citep{Acciari2011, Giordano2012} and thus provides a new opportunity for
studying particle acceleration in SNRs and high energy non-thermal radiation.
Taking the multi-band spectrum into account, it has been shown that the hadronic process,
with energy conversion efficiency of 10--15\% for the current stage, naturally explains
the GeV-TeV \gray\ emission \citep{Giordano2012,Tang2011,Morlino2012}, although
it is argued that a two-zone leptonic model can also work \citep{Atoyan2012}.
In the hadronic scenario, the target gas is always suggested to be a low-density ambient
interstellar medium ($n_{\rm H} \sim 0.2$--$0.3 \cm^{-3}$), which is quite different
from the case of the aforesaid SNRs.

Although it is suggested that Tycho is a naked Ia SNR without any dense
clouds \citep{Tian2011}, there seem to be signs and evidence in multiband
showing the presence of dense clouds from the north to the east (see a
summary in \S\ref{sec:obs}), which could be the target for bombardment of
the accelerated protons and would thus reduce the needed converted energy.
While, theoretically, the acceleration efficiency in SNRs is not well determined,
very efficient acceleration process has been suggested \citep[e.g.,][]{Helder2009},
and some self-consistent models predict that the converted energy fraction
can be up to $\sim 60\%$ -- 80\% in the whole lifetime of an SNR \citep[e.g.,]
[]{Berezhko1997, Kang2006}.
However, constraints from some \gray\ bright young Galactic SNRs hint that
the fraction would not be greater than 10\% at Tycho's age but could reach
10--20\% in subsequent 1--2 kyr, such as the case of RX~J1713.7-3946 \citep[e.g.,][]
{Lee2012}. The similarly young SNR Cas~A has only transferred a minor
fraction ($ \la2\% $) of the total kinetic energy to accelerated particles
\citep{Abdo2010, Araya2010}.
Moreover, under the Bohm limit assumption, the conversion factor no more than 1.3 $\%$
is predicted in kinetic models for a SNR at the Tycho's age, evolving in a pre-shock 
medium of density $\sim0.3\cm^{-3}$,
although its total conversion fraction can be more than 60$\%$ \citep[e.g.,][]{Berezhko1997}.
Therefore, the fraction of order $\sim1\%$ at Tycho's age deserves serious consideration.

Here we present an alternative hadronic scenario with
the modest energy conversion fraction of order $\sim1\%$ for the present Tycho SNR.
As shown below, with such a reduced energy fraction, not only the normal number
ratio between the accelerated electrons and protons is obtained, but also the gas
density in the region of hadronic process will be increased by one order of magnitude,
consistent with the presence of (at least local) dense medium, e.g, molecular clouds (MCs). 
If this is true, the Tycho's SNR will be the only Type Ia SNR associated with MCs
among the currently known interacting SNRs \citep{Jiang2010}.

\section{Hadronic Gamma-ray emission}\label{sect:cal}

We assume that the spectrum of the shock-accelerated particles (electrons and 
protons) obeys a power law with an energy cutoff:
\begin{equation}
 dN_{i}/dE_{i} = A_{i}E_{i}^{-\alpha_{i}}\times \exp (-E_{i}/E_{i,\rm cut})
\end{equation}
where $i={\rm e,p}$, $ E_{i} $ is the particle kinetic energy, $\alpha_{i}$ is the 
spectral index and is fixed as $\alpha_{\rm e,p} \simeq 2.3$ according to the 
afore-mentioned radio index and the charge-independent acceleration process,
$E_{i,\rm cut}$ is the cut-off energy and, for protons, can be taken as 
$ E_{\rm p,max} $ $\simeq$ (37 TeV)$(B_{\rm d}/100\uG)^{2} $ \citep{Parizot2006}
while $E_{\rm e,cut}$ is confined by radiative limit, and
$A_{i}$ is normalization factor and can be determined from
$ \eta E_{\rm SN} = \int^{\infty}_{m_{\rm e}c^{2}} A_{\rm e}E_{\rm e}^{-\alpha_{\rm e}}
\times \exp (-E_{\rm e}/E_{\rm e,cut}) E_{\rm e}dE_{\rm e} + 
\int^{\infty}_{m_{\rm p}c^{2}}A_{\rm p}E_{\rm p}^{-\alpha_{\rm p}}
\times \exp (-E_{\rm p}/E_{\rm p,cut}) E_{\rm p}dE_{\rm p} $
combined with adjustable parameter, \kep = $ A_{\rm e}/A_{\rm p} $, the
number ratio between the accelerated electrons and protons at a
given energy. Here the energy conversion efficiency $\eta \sim 1\%$
is adopted, and $E_{\rm SN}$ is the explosion energy of Tycho.
According to the estimate from the expansion rate of the X-ray
emitting ejecta \citep{Hughes2000}, we will adopt
$E_{\rm SN}=6\times 10^{50}\du^{2} \erg$, where $\du$ is the
distance to the SNR scaled with $2.5\kpc$, in view of the physical
contact of Tycho with interstellar dense clouds (see \S3).

For the \gray\ emission, we consider three potential radiation processes: 
inverse-Compton (IC) scattering on the cosmic microwave background 
by relativistic electrons, nonthermal bremsstrahlung by relativistic electrons
\citep[using the differential cross sections given in][]{Baring1999},
and $\pi^{0}$-decay \grays\ resulting from p--p interaction \citep[using the
parametrization therein]{Kelner2006}.

The downstream magnetic field strength is estimated by X-ray measurements 
and is listed as a function of the distance to this SNR
\citep[Table~5 therein]{Cassam2007}.
Since in this letter we favor an association of Tycho with dense clouds at
a distance of 2.5~kpc (see \S3.3), the field strength can be estimated 
as $B_{\rm d} \simeq 320 \uG$ by interpolation.
This number is also consistent with the estimates (around $300 \uG$) given 
by other methods \citep{Volk2005, Morlino2012}.

\begin{figure}
\centerline { {\hfil\hfill
\epsfig{figure=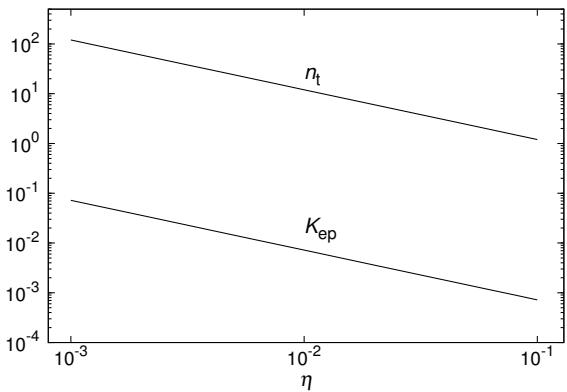,height=2.2in,angle=0}\hfil} }
\caption{Dependence of the parameters, \kep\ and $n_{\rm t}$, on the energy
conversion efficiency $\eta$.}\label{fig:rel}
\end{figure}

From the radio and X-ray synchrotron spectral data and the assumed energy 
conversion efficiency, we get $ E_{\rm e,cut}\approx 5.6 \TeV$ and 
$K_{\rm ep} \approx 0.7 \times 10^{-2} $.
Actually, the product $\eta K_{\rm ep}$ is a constant, with the synchrotron
flux given (shown in Figure~{\ref{fig:rel}}); by contrast,
the cases using an efficiency $\sim10\%$ correspond to a \kep\ value
$\la 10^{-3} $ \citep[e.g.,][]{Morlino2012, Giordano2012}.
Recently a lower limit for the electron to proton ratio, \kep\ $\ga$ $10^{-3}$,
is implied by the radio observations of SNRs in nearby galaxies \citep{Katz2008}.
Actually, \kep\ of order $\sim10^{-2}$ is commonly favoured based on the measured
electron to proton ratio in the CRs
under the assumption that SNRs are the main source of proton and electron CRs
\citep[see e.g.,][]{Longair1994, Katz2008}.
 
\begin{figure}
\centerline { {\hfil\hfil
\epsfig{figure=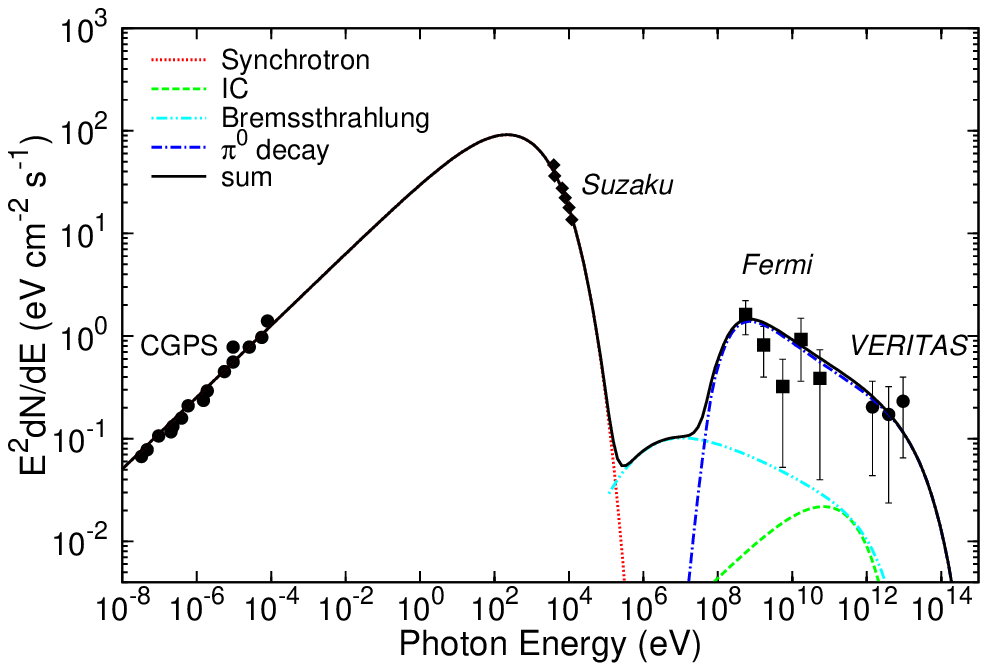,height=2.2in,angle=0}\hfil\hfil } }
\caption
{Broadband SED of Tycho's SNR with the observed data in radio \citep{Kothes2006},
X-rays \citep{Tamagawa2009} and \grays\ (\Fermi : \citealt{Giordano2012};
\veritas : \citealt{Acciari2011}).}\label{fig:sed}
\end{figure}

By fitting \gray\ spectrum, which is here considered to be due to hadronic
emission (see Figure \ref{fig:sed}), and approximating the total energy of
the relativistic protons to be $\eta E_{\rm SN}$, we obtain the product
of $\eta$ and the average density (averaged over the entire shock surface)
of the target protons (with which the energetic particles interact), 
$\nt$, and this product is shown as a $\nt$ -- $\eta$ line in
Figure~\ref{fig:rel}. We have $ \nt \sim 12 \cm^{-3}$ for $\eta \sim 1\%$.
The target protons can be located both upstream and downstream,
since the accelerated particles move back and forth. Downstream, the average
compression factor is estimated to be around 3 \citep{Giordano2012}. Thus the
average preshock target proton density, $n_{0}$, is in the range $\sim$ 4--$12\cm^{-3}$.
This density value is much larger than the gas density ($\nH\sim0.3\cm^{-3}$),
over an order of magnitude, estimated from the previous optical measurement
for the northeast (NE) filament \citep{Kirshner1987} and the \Chandra\ X-ray
observations \citep{Cassam2007, Katsuda2010}.
In \S{\ref{sec:shock-MC}}, we will explain such discrepancy in density for the SNR
shock (as in the NE) expanding in an environment with MCs.

In the above calculation, the diffusive process of the shock accelerated CRs is
not considered. Actually, the typical diffusion radius of CRs around SNRs is
$R_{\rm dif}=2\sqrt{D(E)t_{\rm dif}}$ $\approx0.76\ (t_{\rm dif}/440\yr)^{0.5}\parsec$
for 10 GeV protons \citep[here the correction factor of slow diffusion around
the SNR, $\chi\sim0.01$, has been adopted, see e.g.,][]
{Fujita2009, Giuliani2010, Gabici2010, Li2012}.
Since the diffusion time scale $t_{\rm dif}$ can not be larger than the
SNR's age 440~yr, the diffusion distance for the protons with energy 10~GeV,
which are responsible for $ \sim $1~GeV \grays,
is smaller than 0.8~pc and significantly smaller than the SNR's radius
$\simeq3\ \du \parsec$.
Therefore, in the young SNR Tycho, the hadronic emission does not primarily
arise from the CR protons which collide nearby molecular gas after a long time
diffusion, as in the cases of several interacting SNRs that are considerably
older \citep[$\ge2$~kyr, see e.g.,][]{Li2012}.

\section{Physical contact with dense clouds and the distance}
\subsection{The evidence of shock-cloud interaction}\label{sec:obs}
In the previous section, it is shown that a dense ambient medium 
($n_{0} \sim$ 4 -- 12 $\cm^{-3}$ on average) is inferred,
much denser than the preshock gas that was previously derived ($\sim0.3\cm^{-3}$).
A dense ambient density ($3\cm^{-3}$) was also suggested by a recent hydrodynamic
simulation \citep{Badenes2007}. Indeed, dense medium (molecular gas in particular)
has been suggested to be present in the surroundings of the SNR from the north to
the east based on multi-band observational evidence.

The first evidence comes from the reduced expansion rate of the SNR in the NE compared
to other peripheral parts. Radio \citep{Reynoso1997} and X-ray \citep{Hughes2000,
Katsuda2010} observations show that, along the periphery of the remnant,
although the northeastern section has the largest radius, its expansion index is
smaller than that in all the remaining parts. 
Such evidence suggests that the northeastern peripheral gas currently is moving the
slowest and being remarkably decelerated and that the gas density in the NE is
significantly higher than that in other parts. It is suggested that the shock front in
the NE has reached the outskirts of dense clouds in the recent past \citep{Reynoso1999}. 
In addition, the fact that this slowest expansion part 
has the flattest spectral index $0.44 \pm 0.02$
\citep{Katz-Stone2000} might be another indication for the interaction with
MCs \citep{Koo2001}. Similar phenomenon also appears in the interacting regions
of other SNRs, such as W28 \citep{Dubner2000} and IC443 \citep{Green1986}.

Secondly, there seem to be some direct signs of the presence of adjacent
dense clouds from the CO line observations.
The \twCO\ ($J =$1--0) observations show that
the MCs at $-68\km\ps \le V_{\rm LSR} \le -58\km\ps$ forms a semi-closed non-uniform
large shell surrounding the SNR from the north to the east and the inner boundary
essentially traces the rim of Tycho \citep*{Lee2004, Cai2009}. 
A high-resolution ($16''$) close-up observation shows that
the $-63.5\km\ps \le V_{\rm LSR} \le -61.5\km\ps$ \twCO\ ($J=$1--0) emitting area
appears to be in contact with the remnant along its northeastern boundary \citep{Lee2004}. Specifically, there is a big molecular clump in the integration map of this
velocity component \citep[see Fig.~1 in][]{Lee2004} at the azimuth corresponding to
the slowest expansion \citep{Reynoso1997}. A similar emitting zone at $\sim-62.5
\km\ps$ is also revealed by the submillimeter observation in the higher level
transition line, \twCO\ ($J=$2--1) \citep*{Xu2011}. The virial mass of the surrounding
MCs derived from the \twCO\ ($J=$1--0) emission is found to be larger than their
gravitational mass, implying that the MCs are being violently disturbed and Tycho's
shock is very likely to be responsible for the disturbance \citep{Cai2009}. Such
disturbance seems to be consistent with the broad \twCO\ ($J=2$--1) lines present in $-68$
to $-55\km\ps$ interval for a few adjacent MCs \citep{Xu2011}.

Thirdly, interaction with dense medium is also favoured by the far- to mid-IR
observations \citep{Ishihara2010, Gomez2012}.
The far-IR emissions (\AKARI : 140 and $160\um$; \herschel : 160, 250, 350 and
$500\um$), which originate from cold dust, are spatially correspondent to the
CO emission in $-68$ to $-53\km\ps$ interval.
The mid-IR emissions (\AKARI : 15, 18, and $24\um$; \herschel : 24 and $70\um$),
which primarily shows warm dust distribution, reveal a 
shell-like emission structure along the remnant boundary except in the west
and southwest and seem to delineate the interface between the remnant and the
MCs. They have been pointed out to arise from the outer edge of the shock
heated cold dust and MCs, although the possibility that a small fraction of
the warm dust was formed in the supernova ejecta was not absolutely ruled out
\citep{Ishihara2010, Gomez2012}.
The mid-IR filter bands, such as \AKARI\ L15 (12.6--$19.4\um$) and L18W 
(13.9--$25.6\um$), may also include the contribution from the rotational lines of 
hydrogen molecules \citep[like S(1) $J=3$--1 at $17.03\um$;][]{Ishihara2010}.
As an alternative, the mid-IR emissions are suggested to come from the swept-up
interstellar dust through interaction with the HI clump in the NE, shown as a
significant absorption feature in the $-52.7\km\ps$, which is also at the 
azimuthal range of the slowest expansion and the faint overlapping $-55\km\ps$
MC in the NW \citep{Gomez2012}.

\subsection{Discussion about the shock in cloudy medium}\label{sec:shock-MC}
The presence of dense clouds along the boundary, especially in the NE, and 
the large target-particle density derived here can be compatible with the 
low density preshock gas that is previously inferred from optical and X-ray
observations. Here we mention two possible cases for the shock-cloud interaction.

The first case is that the blast shock may be flanking a cloud (e.g, the
slowed-expansion portion in the NE) and the shocked gas outside the cloud
(but along the line of sight) has a density of order $0.1\cm^{-3}$ as observed
in X-rays and optical. The solid angle fraction of the whole remnant covered
by the interacting region may roughly be $\fo 
\sim 0.07$, where a subtended angle $\sim 60^\circ$ is adopted 
from the slowed-expansion portion \citep{Reynoso1997}.
The molecular gas surrounding Tycho may have a density $\nMC$ as high as
$\sim10^{2} \cm^{-3}$ \citep{Tian2011}. Thus the average preshock proton density is
$n_0=\fo (2n_{\rm MC})\sim14(\nMC/100\cm^{-3})\cm^{-3}$, which is very similar to
the value we obtain above (considering the uncertainties in the used
parameters).

The alternative case is that the shock may be expanding into a clumpy medium and the
low density corresponds to the interclump medium that fills in most of the 
volume. For instance, blast shocks are found to propagate into the interclump
gas of densities of order $0.1\cm^{-3}$ in the southeastern edge of Kes~69
\citep{Zhou2009} and in the northeastern edge of Kes~78 \citep{Zhou2011},
where the shocks are interacting with MCs, as derived from the X-ray analyses;
in W28, W44, and 3C 391, where blast waves are impacting MCs, the preshock
density of the interclump gas is also $<1\cm^{-3}$ as inferred from the IR
spectroscopic study \citep{Reach2000}.
If $n_0\sim4$--$12\cm^{-3}$ derived above from the hadronic emission
is used and the clump filling factor in the dense cloud is denoted as $\fc$
(which may be of order 0.1 or smaller),
then the solid angle fraction covered by the cloud(s) is
$\fo=n_0/(2\nMC\fc)\sim0.2$--$0.6(\nMC/100\cm^{-3})^{-1}(\fc/0.1)^{-1}$.
Such a solid angle seems to be consistent with the distribution of semi-closed
dense gas surrounding the SNR as suggested by the CO line and IR observations
(\S\ref{sec:obs}).

We note that there is not yet report of detection of optical emission from
the shocked clouds. One of the possibilities may be that the shock collision
with the clouds is a very recent event, e.g., within 50 yr \citep{Reynoso1999},
which is shorter than the radiative cooling time $\sim300\yr$ \citep[using Eq.(3)
in][with the velocity of the blast shock $0.3''\yr^{-1}$ at 2.5~kpc 
or $3500\km\ps$ adopted]{Sgro1975}. 
Little thermal X-rays are either detected from the shocked clouds. A similar
case is seen in RX~J1713.7-3946, another young SNR interacting with MCs
\citep{Moriguchi2005}.
It is pointed out that the substantial suppression of the thermal X-ray emission
could result from the very low intercloud gas density and the stall of the 
transmitted shock in the clouds \citep{Zirakashvili2010, Inoue2012}. 

\subsection{Distance to Tycho}
The SNR-cloud association, in combination with the line-of-sight HI absorption
\citep{Tian2011}, can be used to estimate the distance to the Tycho's SNR,
which has long been in debate due to its special location in our Galaxy \citep[see a
brief summary in Figure~6 of][]{Hayato2010}. 
The LSR velocity $\sim-62\km\ps$ of the semi-closed molecular shell (including
the northeastern MCs) allows two alternatives for the distance, 4.6~kpc and
2.5~kpc, while the LSR velocity $\sim-55\km\ps$ of the faint northwestern MC
allows 4.0~kpc and 2.5~kpc. The latter estimates are because of the deep ``dip"
at 2.5~kpc, where the spiral shock front is located, in the rotation curve for
the longitude $120^{\circ}$ \citep[\citealt{Tian2011};
using the \citeauthor{Foster2006}' \citeyearpar{Foster2006} model; 
also see Figure~14 of][]{Schwarz1995} and the $b$-dependent
behaviour of the velocity field in the spiral shock region, in which the systematic
velocity of the MCs could be as low as $\la-60\km\s^{-1}$ \citep{Lee2004}.
Due to the absence of the absorption feature for the HI gas at $-41$ to $-46\km\ps$,
Tycho should be in front of this gas, as \citet{Tian2011} constraint it to be
nearer than $3\kpc$, and thus the alternative distances of $4.6\kpc$ and $4.0\kpc$ 
are excluded. 
In the scenario of the association with the $\sim-62\km\s^{-1}$ velocity component,
the HI gas $\sim-60\km\s^{-1}$ \citep[lack of absorption feature,][]{Tian2011} is
behind the SNR, and the $-47$ to $-53\km\ps$ HI absorbing gas is in the front, but
all of them are closely around the spiral shock at the``dip", at the distance of
2.5~kpc. In the scenario of the association between the remnant and the $\sim-55\km\s^{-1}$
MC and the $-52.7\km\ps$ HI cloud (see \S{\ref{sec:obs}}), the situation
of HI absorption and SNR location
is similar. Tycho is located in the high density region just behind the spiral
shock \citep{Lee2004} and is now encountering the inhomogeneous outskirts of the
$\sim-62\km\s^{-1}$ or/and $\sim-55\km\ps$ MCs.

\section{conclusion}
Tycho is one of nearly a dozen Galactic SNRs which are suggested to radiate hadronic
\gray\ emission. It is noted, however, that it is the only one in which the hadronic 
emission is proposed to arise from the interaction with low-density ($\sim0.3\cm^{-3}$)
ambient medium. An alternative explanation that the \grays\ originate from hadronic
process with fewer energetic protons but denser target gas is presented here.
With a conversion efficiency of order 1\% for this young historical SNR,
a normal electron-proton ratio (of order
$10^{-2}$) is derived from the radio and X-ray synchrotron spectra and an average
ambient density that is at least one-order-of-magnitude higher is further derived
from the hadronic \gray\ flux. 
This result is consistent with the multi-band observational evidence of the presence of the
surrounding dense medium from the north to the east. The interaction scenario,
combined with the HI absorption data, is used to constrain the long controversial
problem of the distance to Tycho and leads to an estimate of 2.5~kpc.

\section*{Acknowledgments}
The anonymous referee is acknowledged for valuable comments and advices, which
have helped to improve the manuscript. We thank Wenwu Tian, Siming Liu, Daisuke
Ishihara and Ping Zhou for helpful advices. 
Y.C.\ acknowledges support from the 973 Program grant 2009CB824800 and NSFC
grants 11233001 and 10725312. X.Z.\ thanks support from the China Postdoctoral
Science Foundation grant 2011M500963.

\bibliographystyle{mn2e}
\bibliography{reference}

\label{lastpage}
\end{document}